\documentclass[conference]{IEEEtran}
\usepackage[numbers]{natbib}
\usepackage{mciteplus}
\usepackage{amsmath,amssymb,amsfonts}
\usepackage{graphicx}
\usepackage{textcomp}
\usepackage{xcolor}
\usepackage{booktabs}
\usepackage{multirow}
\usepackage{array}
\usepackage{placeins}
\usepackage{float}
\usepackage{tikz}
\usetikzlibrary{shapes.geometric, arrows.meta, positioning, calc, fit, backgrounds, shadows.blur}

\graphicspath{{figures/}}
\def\BibTeX{{\rm B\kern-.05em{\sc i\kern-.025em b}\kern-.08em
    T\kern-.1667em\lower.7ex\hbox{E}\kern-.125emX}}

\begin{document}

\title{Toward Localizing and Repairing Bias in Transformer Attention Heads}

\author{\IEEEauthorblockN{Sigma Jahan}
\IEEEauthorblockA{Dalhousie University, Canada\\
\texttt{sigma.jahan@dal.ca}}
}

\maketitle

\begin{abstract}
Transformer language models are increasingly used as software components, yet biased outputs remain difficult to localize and repair inside the model. Existing fairness testing and repair methods largely operate at the input-output or retraining level, while recent work suggests that bias-related behavior can concentrate in a small set of attention heads. This paper studies whether attention heads can be localized and repaired through a targeted inference-time intervention. We introduce \textsc{ROBIN}, a white-box head-level fairness debugging method that ranks attention heads using sensitivity to fairness probes and removes a small bias subspace from selected head outputs. In a four-model pilot study, \textsc{ROBIN} reduces the measured WinoBias gap across all models while preserving language-modeling quality better than whole-head zeroing. These preliminary results suggest that head-level bias repair should consider not only which heads are selected, but also how selected heads are modified.
\end{abstract}

\begin{IEEEkeywords}
bias, fairness, debugging, attention heads
\end{IEEEkeywords}

\section{Introduction}

Real-world deployments of machine-learning software systems have produced documented unfair outcomes. Amazon withdrew an internal resume-screening system after engineers found that it ranked applicants lower when their resumes mentioned women's colleges or the word ``women's''~\cite{dastin2018amazon}. For ML-enabled software maintenance, this case exposes a debugging gap. Developers may reproduce a fairness failure at the system boundary yet still lack an internal location or a bounded repair. Similar concerns appear across natural language processing systems, where fairness failures are visible in outputs but difficult to trace to internal model behavior~\cite{bender2021dangers}.

Software engineering research has developed two complementary directions for addressing fairness bugs. The first direction focuses on testing software for discriminatory behavior. \citet{galhotra2017fairness} introduced fairness testing as a quality property of the software itself, and \citet{udeshi2018automated} developed directed test generation to expose individual discriminatory inputs. Adjacent work on testing and debugging deep neural networks more broadly, including concolic testing~\cite{sun2018concolic} and state-differential model debugging~\cite{ma2018mode}, treats neural models as software artifacts amenable to traditional analysis. Surveys synthesize this work and find that fairness defects are difficult to characterize because their causes depend on data, model behavior, protected attributes, and deployment context~\cite{chakraborty2021bias,chen2024fairness}.

The second direction repairs models after a failure. \citet{chakraborty2020fairway} reweight or relabel training data to reduce bias before deployment, and \citet{zhang2021ignorance} study how repair interacts with accuracy loss. These methods help practitioners detect and mitigate unfair behavior, but they operate from outside the model. They reveal unequal treatment or retrain the model without identifying which internal structures contribute to an inference-time fairness failure. A white-box debugging method could complement them by localizing candidate components and applying a bounded change without retraining the full model.

Recent work suggests that bias-related behavior in masked language models can concentrate in a small set of attention heads~\cite{yang2023bias}. This finding narrows the localization search but does not determine how a selected head should be modified. Head-attribution methods rank heads using gradients or activations~\cite{michel2019sixteen,voita2019analyzing}. Whole-head zeroing then provides a simple intervention, but it assumes that a selected head primarily encodes bias even though one head may also support general linguistic behavior~\cite{clark2019what,voita2019analyzing}. We instead seek an operator that retains output components orthogonal to the estimated bias. Our design combines empirical-Fisher importance scoring~\cite{kirkpatrick2017overcoming} with multi-dimensional projection derived from bias-direction and null-space methods~\cite{bolukbasi2016man,ravfogel2020null}. This combination separates which heads to inspect from how to modify them.

A selected attention head may contribute to both the measured fairness gap and ordinary language modeling behavior, so zeroing the whole head can remove useful information along with the biased response. In this paper, we introduce \textbf{ROBIN} (\textbf{R}epair \textbf{O}f \textbf{B}ias via \textbf{IN}tervention), a white-box method for head-level fairness debugging in transformer language models. \textsc{ROBIN} ranks attention heads by their sensitivity to fairness probes and removes a small bias subspace from the top-$k$ head outputs at inference. We evaluate \textsc{ROBIN} in a pilot study on DistilBERT, BERT, DistilGPT-2, and GPT-2. At $k=20$, \textsc{ROBIN} reduces the WinoBias gap on all four models with at most a $7.2$\% increase in language-modeling loss.

This paper makes the following contributions:
\begin{itemize}
    \item We frame head-level fairness mitigation as a white-box software debugging task that separates head localization from component-level modification.
    \item We propose \textsc{ROBIN}, a method that identifies bias-sensitive attention heads and removes only the biased part of each head at inference.
    \item We provide initial evidence on four pretrained transformer models that \textsc{ROBIN} reduces the measured WinoBias gap while limiting the language-modeling loss increase to $7.2$\%.
\end{itemize}

\section{Methodology}
\label{sec:robin}

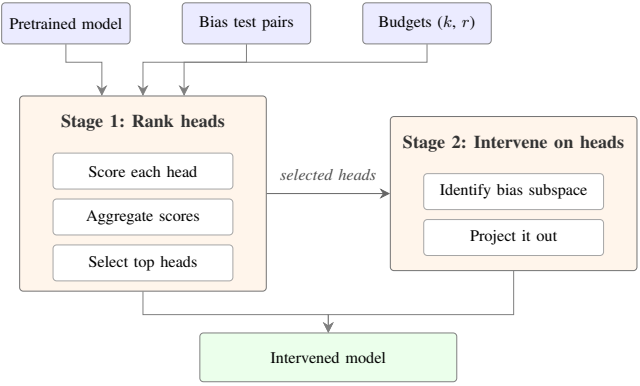
\begin{figure}[!htbp]
\centering
\resizebox{0.95\columnwidth}{!}{%
\begin{tikzpicture}[
    >={Stealth[length=2mm, width=2mm]},
    inputbox/.style={
        rectangle, rounded corners=2pt,
        draw=black!50, line width=0.4pt,
        fill=blue!8,
        minimum width=2.5cm, minimum height=0.80cm,
        align=center, font=\small
    },
    stagebox/.style={
        rectangle, rounded corners=2pt,
        draw=black!55, line width=0.4pt,
        fill=orange!8,
        minimum width=4.8cm, minimum height=3.8cm,
        align=center
    },
    stagelabel/.style={
        font=\normalsize\bfseries, text=black!80
    },
    stepbox/.style={
        rectangle, rounded corners=2pt,
        draw=black!35, line width=0.4pt,
        fill=white,
        minimum width=3.5cm, minimum height=0.7cm,
        align=center, font=\small
    },
    outputbox/.style={
        rectangle, rounded corners=2pt,
        draw=black!50, line width=0.4pt,
        fill=green!8,
        minimum width=5.0cm, minimum height=0.95cm,
        align=center, font=\small
    },
    arr/.style={->, draw=black!50, line width=0.4pt},
    midarr/.style={->, draw=black!60, line width=0.5pt}
]

\node[inputbox] (model) {Pretrained model};
\node[inputbox, right=1.0cm of model] (probes) {Bias test pairs};
\node[inputbox, right=1.0cm of probes] (budget) {Budgets ($k$, $r$)};

\node[stagebox, below=1.0cm of model, xshift=1.5cm] (stage1) {};
\node[stagelabel, anchor=north] (s1t) at ([yshift=-0.25cm]stage1.north)
    {Stage 1: Rank heads};
\node[stepbox, anchor=north] (s1a) at ([yshift=-0.30cm]s1t.south)
    {Score each head};
\node[stepbox, anchor=north] (s1b) at ([yshift=-0.18cm]s1a.south)
    {Aggregate scores};
\node[stepbox, anchor=north] (s1c) at ([yshift=-0.18cm]s1b.south)
    {Select top heads};

\node[stagebox, right=2.4cm of stage1, minimum height=3.0cm] (stage2) {};
\node[stagelabel, anchor=north] (s2t) at ([yshift=-0.25cm]stage2.north)
    {Stage 2: Intervene on heads};
\node[stepbox, anchor=north] (s2a) at ([yshift=-0.30cm]s2t.south)
    {Identify bias subspace};
\node[stepbox, anchor=north] (s2b) at ([yshift=-0.18cm]s2a.south)
    {Project it out};

\node[outputbox,
      below=1.0cm of $(stage1.south)!0.5!(stage2.south)$]
      (out) {Intervened model};

\draw[arr] (model.south)  -- ++(0,-0.35) -| ([xshift=-0.8cm]stage1.north);
\draw[arr] (probes.south) -- ++(0,-0.35) -| (stage1.north);
\draw[arr] (budget.south) -- ++(0,-0.35) -| ([xshift=0.8cm]stage1.north);

\draw[midarr] (stage1.east) --
    node[above=4pt, font=\small\itshape, text=black!70] {selected heads}
    (stage2.west);

\draw[arr] (stage1.south) |- ([yshift=0.35cm]out.north) -- (out.north);
\draw[arr] (stage2.south) |- ([yshift=0.35cm]out.north) -- (out.north);

\end{tikzpicture}%
}
\caption{\textsc{ROBIN} workflow}
\label{fig:workflow}
\end{figure}

\textsc{ROBIN} assumes white-box access to a trained transformer language model~\cite{vaswani2017attention} $f_\theta$ and to its tokenizer. The fairness probes are a set of $N$ text pairs $(x_a^{(i)}, x_b^{(i)})$ that differ only in the \textit{protected attribute} (the demographic dimension along which we measure unfairness, e.g., gender or race). For WinoBias the difference is a gender pronoun (he/she), and for StereoSet it is a sentence-level change between a stereotype and an anti-stereotype version. Given these inputs and two user-specified budgets, $k$ (how many heads to modify) and $r$ (how many bias directions to remove from each chosen head), \textsc{ROBIN} operates in two stages (Fig.~\ref{fig:workflow}). \textit{Stage 1} ranks every attention head by its sensitivity to the difference between $x_a$ and $x_b$. \textit{Stage 2} modifies the top $k$ heads at inference.

\textit{Per-pair score.} The method assigns each text a confidence score computed only at the positions where the paired texts differ. We denote those positions by $D$. For a masked language model, we mask the changed positions and define
\begin{equation}
p(\text{text}) = \exp\!\left(-\tfrac{1}{|D|}\sum_{t \in D} \mathcal{L}_t\right)
\label{eq:p_mlm}
\end{equation}
where $\mathcal{L}_t$ is the negative log-likelihood of the original token at position $t$. For a causal language model, we instead run the model unmasked and use the probability of the original token at each changed position,
\begin{equation}
p(\text{text}) = \exp\!\left(\tfrac{1}{|D|}\sum_{t \in D} \log P_\theta(w_t \mid w_{<t})\right)
\label{eq:p_clm}
\end{equation}
In both cases $p(\text{text}) \in [0, 1]$ measures the model's confidence on the changed positions.

\textit{Bias metrics.}
\label{sec:metrics}
We report two bias metrics built from $p$. The \textit{WinoBias gap} is the mean absolute difference of $p$ between the two sides of a pair, where lower indicates a smaller gap.
\begin{equation}
\text{Gap} = \frac{100}{N}\sum_{i=1}^{N} \left| p(x_a^{(i)}) - p(x_b^{(i)}) \right|
\label{eq:gap}
\end{equation}
The \textit{StereoSet stereotype score} (SS) is the percentage of pairs where the model prefers the stereotype sentence over the anti-stereotype. A value of $50$ means no preference. A value above $50$ indicates a stereotype preference under this metric.
\begin{equation}
\text{SS} = \frac{100}{N}\sum_{i=1}^{N} \mathbb{1}\!\left[\,p(\text{stereo}^{(i)}) > p(\text{anti}^{(i)})\,\right]
\label{eq:ss}
\end{equation}

\textbf{1. Rank attention heads.} Stage 1 turns the probe pairs into a list of attention heads, from most to least bias-sensitive.

\textit{1a. Per-pair gradient.} For each pair we run the model twice, once on $x_a$ and once on $x_b$, with loss $\mathcal{L} = -\log p(\text{text})$. We record the size of the loss gradient with respect to the attention weights inside each head,
\begin{equation}
g_h^{(i)} \;=\; \left\| \nabla_{A_h}\, \mathcal{L} \right\|_2
\label{eq:g_h}
\end{equation}
where $A_h$ is the attention-weight matrix of head $h$ on side $i$. A larger $g_h^{(i)}$ indicates greater local sensitivity of the loss to that head's attention weights. Each pair gives us two such numbers per head.

\textit{1b. Gradient score.} We average the gradient sizes from Step 1a over all $2N$ runs (two per pair) to get one number per head,
\begin{equation}
G_h \;=\; \frac{1}{2N}\sum_{i=1}^{2N} g_h^{(i)}
\label{eq:G_h}
\end{equation}
A head with a large $G_h$ has high average sensitivity to the gender change.

\textit{1c. Squared-gradient score.} Alongside the simple average, we also compute the average of the \emph{squared} gradient sizes,
\begin{equation}
C_h \;=\; \frac{1}{2N}\sum_{i=1}^{2N} \left(g_h^{(i)}\right)^{\!2}
\label{eq:C_h}
\end{equation}
Squaring gives more weight to heads with a few large per-pair responses. The mean-gradient score $G_h$ favors heads that respond consistently across many probe pairs, whereas $C_h$ favors heads whose response concentrates on a smaller subset. We use both since bias-responsive heads can take either form. Some may react broadly across pairs, while others may react sharply on a few and stay quiet elsewhere~\cite{yang2023bias}. The squared-gradient form follows empirical-Fisher importance scoring used in Elastic Weight Consolidation~\cite{kirkpatrick2017overcoming}, applied at the attention-head level following head-importance scoring in prior pruning work~\cite{michel2019sixteen}.

\textit{1d. Combine the scores.} The $G_h$ and $C_h$ values are on different scales, so we put each on a common scale by subtracting its mean and dividing by its standard deviation. Writing this rescaling as $z(\cdot)$, the combined \textsc{ROBIN} score is
\begin{equation}
F_h \;=\; z(G_h) + z(C_h)
\label{eq:F_h}
\end{equation}
The combined score therefore ranks heads highly when they show both broad sensitivity across probe pairs and concentrated high sensitivity on some pairs. We weight the standardized terms equally to avoid tuning a model-specific coefficient in this pilot study.

\begin{figure}[!htbp]
\centering
\includegraphics[width=0.68\columnwidth]{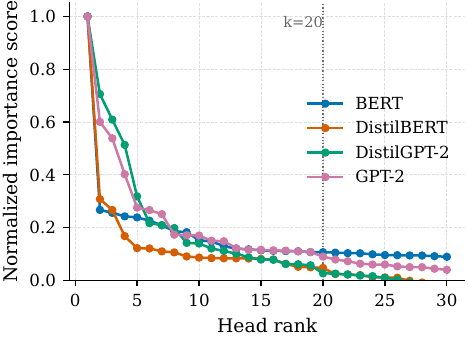}
\caption{Top-$30$ \textsc{ROBIN} importance scores, per-model normalized}
\label{fig:score_concentration}
\end{figure}

\textit{1e. Pick the top $k$ heads.} We rank heads by $F_h$ from largest to smallest and keep the top $k$ as the chosen set $\mathcal{S}_k$. In our experiments we try $k \in \{1, 2, 3, 5, 8, 10, 15, 20\}$. The importance score drops sharply within the first few ranks on every model we evaluate and flattens well before rank $30$ (Fig.~\ref{fig:score_concentration}), which is the empirical reason we restrict $k$ to small values rather than searching higher budgets.

\textbf{2. Intervene on chosen heads.} Stage 2 takes the $k$ heads from Stage 1 and modifies their outputs at inference.

\textit{2a. Find the bias subspace.} For each head $h \in \mathcal{S}_k$, we estimate a bias subspace, which we define as the directions in the head output that change most when the protected attribute is altered. We take up to $400$ WinoBias pairs and, for each pair, average the head's output on the changed-token positions to obtain one vector for $x_a$ and one for $x_b$. Their difference $d^{(i)} = a^{(i)} - b^{(i)} \in \mathbb{R}^{d_h}$, where $d_h$ is the head's output dimension, is one example of how the head's output shifts under this change. Stacking these difference vectors as rows gives a matrix $D_h$ with up to $400$ rows. If the bias signal followed one direction, the $d^{(i)}$ vectors would be nearly aligned. In practice, they occupy several directions, so we keep the $r$ directions with the largest singular values. These are the top $r$ right singular vectors of $D_h$,
\begin{equation}
V_h \;=\; \mathrm{TopRightSVD}_r(D_h) \;\in\; \mathbb{R}^{r \times d_h}
\label{eq:V_h}
\end{equation}
We use $r = 4$ throughout. Setting $r = 1$ recovers the single-direction estimate of \citet{bolukbasi2016man}.

\textit{2b. Remove the bias subspace.} When the model runs on new text, each chosen head computes its output $x_h$ as usual, and we then subtract the part of $x_h$ that lies along the bias subspace $V_h$,
\begin{equation}
x'_h \;=\; x_h - V_h^{\!\top} V_h\, x_h
\label{eq:proj}
\end{equation}
The term $V_h^{\!\top} V_h\, x_h$ is the projection of $x_h$ onto the bias subspace. Removing this projection keeps the remaining part of the head output. For batch size $B$ and sequence length $T$, the added cost is $O(BT d_h r)$ per chosen head.

\textit{2c. Subspace removal vs. head ablation.} Attention heads in transformer language models can support multiple functions~\cite{voita2019analyzing,clark2019what}. A head that responds strongly to gendered pronouns may also support general linguistic behavior, including this kind of pronoun-tracking. Zeroing the whole head removes both at once. Subtracting only the part of $x_h$ that lies along $V_h$ removes the bias response and keeps the rest. As Section~\ref{sec:results} reports, this distinction matters empirically.

\section{Experimental Design}

\textbf{Models.} We evaluate \textsc{ROBIN} on four publicly available pretrained transformer language models that span both the encoder and decoder families and both base-size and distilled variants. The encoders are BERT~\cite{devlin2019bert} and DistilBERT~\cite{sanh2019distilbert}, which we score as masked language models. The decoders are GPT-2~\cite{radford2019language} and DistilGPT-2, which we score as causal language models. The two base-size models have $144$ attention heads each ($12$ layers $\times\, 12$ heads), and the two distilled models have $72$ ($6$ layers $\times\, 12$ heads), a difference that becomes relevant when we discuss inference cost in Section~\ref{sec:results}.

\textbf{Datasets.} We use two paired bias benchmarks and one general corpus. \textit{WinoBias}~\cite{zhao2018gender} provides sentences in which the model must resolve a pronoun (he or she) to one of two professions, set up so that the stereotype-aligned resolution differs from the anti-stereotype one. We form pairs by replacing every gendered pronoun with its opposite, giving $1{,}584$ pairs after filtering pairs unchanged by the replacement. \textit{StereoSet}~\cite{nadeem2021stereoset} provides intrasentence stereotype-vs-anti-stereotype pairs across gender, race, profession, and religion, giving $2{,}106$ pairs. \textit{WikiText-2}~\cite{merity2017regularizing} provides general text on which we measure language-modeling loss. WinoBias drives the head ranking and the main bias result. StereoSet serves as a secondary fairness check across protected categories.

\textbf{Compared Methods.} We compare three head-importance scores under the same evaluation setup. All three derive from the per-pair gradient norm $g_h^{(i)}$ defined in Eq.~(\ref{eq:g_h}). They differ in how we aggregate the per-pair scores and in how we modify selected heads at inference.

\textit{- Gradient.} The per-head mean of $g_h^{(i)}$ across all pair-sides, $G_h$ in Eq.~(\ref{eq:G_h}). At inference, we zero the top $k$ heads by setting their slice of the attention output projection input to zero before the projection runs.

\textit{- SquaredGrad.} The per-head mean of $(g_h^{(i)})^2$, $C_h$ in Eq.~(\ref{eq:C_h}). We zero the top $k$ heads in the same manner as for Gradient.

\textit{- \textsc{ROBIN}.} The standardized sum $F_h = z(G_h) + z(C_h)$ in Eq.~(\ref{eq:F_h}). We do not zero the top $k$ heads. Instead, for each selected head we estimate a four-dimensional bias subspace $V_h$ (Eq.~(\ref{eq:V_h})) and replace the head output $x_h$ with $x_h - V_h^\top V_h x_h$ (Eq.~(\ref{eq:proj})) at inference. Projecting out a learned bias subspace at the representation level has precedent in sentence-level debiasing~\cite{liang2020towards} and in iterative null-space methods~\cite{ravfogel2020null}, but those operate on whole-model outputs rather than on a small number of selected heads.

This setup separates the choice of head-importance score from the choice of inference-time modification. We vary $k$ over $\{1, 2, 3, 5, 8, 10, 15, 20\}$, fix $r = 4$ across all models (chosen from a BERT pilot comparison), and report means over three seeds. Across seeds, we vary the subset of probe pairs used to estimate the bias subspace and the masking pattern applied when scoring encoder quality.

\textbf{Metrics.} We report four metrics, two for fairness, one for model performance, and one for inference cost.

\textit{- WinoBias gap (Gap).} Lower is better for this metric. An unmodified model with Gap close to $0$ assigns similar scores to the two pronoun forms (see Eq.~\ref{eq:gap}).

\textit{- StereoSet stereotype score (SS).} A value of $50$ indicates no stereotype preference. We report SS as secondary evidence, since it is less stable than Gap in our experiments and we do not base our main fairness claim on it (see Eq.~\ref{eq:ss}).

\textit{- Language-modeling loss.} We measure model performance on WikiText-2 using a single metric we refer to as \textit{language-modeling loss}, with lower values indicating better performance. The exact formula depends on the model family. For decoders (GPT-2, DistilGPT-2), it is the standard token-level perplexity over length-$128$ chunks, defined as $\exp$ of the mean negative log-likelihood. Encoders (BERT, DistilBERT) do not define an autoregressive likelihood, so we instead report a masked-token pseudo-perplexity at $15$\% masking, following~\citet{salazar2020masked}. We use one umbrella term for readability, but encoder and decoder values should be compared only within model family, not across.

\textit{- Inference latency.} The average time the model takes to process one batch of input on a single GPU. We report it as the percentage overhead each method adds compared to the unmodified model. Details of the hardware and batch setup are in the replication package.

\section{Results}
\label{sec:results}
\noindent
{\textbf{RQ1. Does \textsc{ROBIN} reduce the WinoBias gap in pretrained transformer language models?}}

\noindent
At $k=20$, \textsc{ROBIN} reduces the WinoBias gap on every model (Table~\ref{tab:main_results}). The largest absolute reduction is on BERT, the model that started most biased ($45.97 \to 19.14$). The other three models show smaller absolute reductions. StereoSet stereotype score moves inconsistently across models and protected categories, so we treat it as secondary and answer RQ1 using WinoBias gap as the primary fairness metric.

\vspace{-1pt}
\begin{table}[!htbp]
\caption{Bias and language-modeling loss at $k = 20$}
\label{tab:main_results}
\centering
\scriptsize
\setlength{\tabcolsep}{3pt}
\textit{(a) Bias.} Gap and SS by model and method\\[1pt]
\begin{tabular}{l rr rr rr rr}
\toprule
& \multicolumn{2}{c}{Baseline} & \multicolumn{2}{c}{Gradient} & \multicolumn{2}{c}{SquaredGrad} & \multicolumn{2}{c}{\textsc{ROBIN}} \\
\cmidrule(lr){2-3} \cmidrule(lr){4-5} \cmidrule(lr){6-7} \cmidrule(lr){8-9}
Model & Gap & SS & Gap & SS & Gap & SS & Gap & SS \\
\midrule
DistilBERT  & 10.78 & 56.32 &  2.15 & 55.46 &  2.15 & 55.46 &  2.51 & 57.03 \\
BERT        & 45.97 & 54.75 & 20.73 & 56.17 & 27.08 & 56.98 & 19.14 & 55.75 \\
DistilGPT-2 & 14.65 & 61.87 &  4.49 & 62.92 &  4.18 & 62.49 &  6.95 & 62.63 \\
GPT-2       & 15.34 & 60.16 & 10.50 & 62.30 & 10.50 & 62.30 &  9.78 & 61.68 \\
\bottomrule
\end{tabular}\\[6pt]
\textit{(b) Language-modeling loss on WikiText-2}\\[1pt]
\begin{tabular}{l rrrr}
\toprule
Model & Baseline & Gradient & SquaredGrad & \textsc{ROBIN} \\
\midrule
DistilBERT  &  13.56 &  25.16 &  25.16 &  14.48 \\
BERT        &  12.11 &  13.43 &  15.93 &  11.98 \\
DistilGPT-2 & 126.23 & 311.88 & 250.00 & 135.29 \\
GPT-2       &  71.35 &  96.31 &  96.31 &  75.59 \\
\bottomrule
\end{tabular}\\[2pt]
{\scriptsize\itshape Note. Lower is better for Gap and language-modeling loss. SS closer to $50$ is better.}
\end{table}

\noindent\textit{\textbf{Finding 1.} At $k=20$, \textsc{ROBIN} reduces the WinoBias gap on all four models and more than halves it on three, with the largest absolute drop on BERT ($45.97 \to 19.14$).}

\begin{figure}[!htbp]
\centering
\begin{minipage}[t]{0.49\linewidth}
\centering
\includegraphics[width=\linewidth]{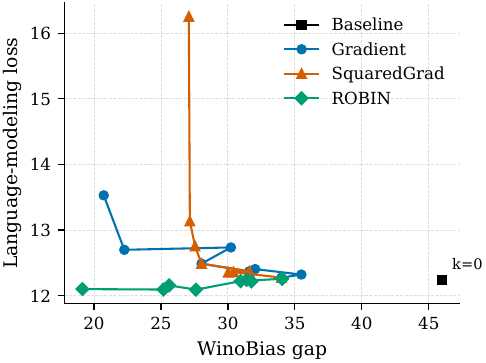}\\[-0.3em]
{\scriptsize (a) BERT}
\end{minipage}\hfill
\begin{minipage}[t]{0.49\linewidth}
\centering
\includegraphics[width=\linewidth]{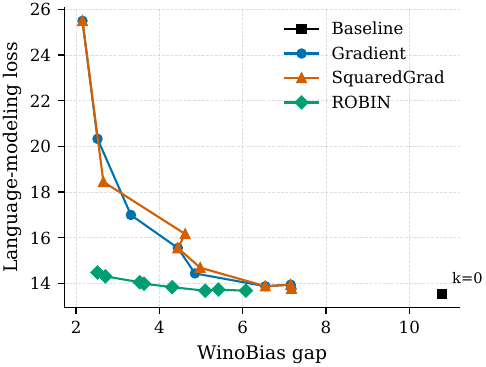}\\[-0.3em]
{\scriptsize (b) DistilBERT}
\end{minipage}
\caption{WinoBias gap vs.\ language-modeling loss on WikiText-2 as $k$ varies}
\label{fig:pareto}
\end{figure}

\begin{figure}[!htbp]
\centering
\includegraphics[width=0.62\columnwidth]{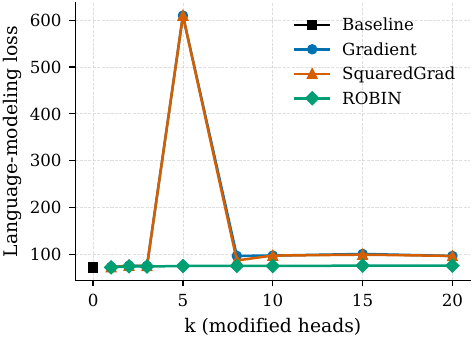}
\caption{GPT-2 language-modeling loss on WikiText-2 as $k$ varies}
\label{fig:gpt2_quality}
\end{figure}

\vspace{1em}
\noindent
{\textbf{RQ2. How does \textsc{ROBIN} trade off WinoBias-gap reduction against language-modeling loss?}}\\
\noindent
As $k$ grows, Gap generally decreases while language-modeling loss rises (Fig.~\ref{fig:pareto}). \textsc{ROBIN} limits this rise compared with whole-head zeroing. On BERT, it reaches the lowest $k=20$ Gap at near-baseline language-modeling loss. On DistilBERT, its Gap is slightly higher than SquaredGrad ($2.51$ versus $2.15$), but the increase in loss is $11$ points smaller ($14.48$ versus $25.16$). On GPT-2, \textsc{ROBIN} stays within a few loss points of baseline across $k$, whereas zeroing ends with a $35$\% increase at $k=20$ (Fig.~\ref{fig:gpt2_quality}). DistilGPT-2 follows the same pattern at a larger absolute scale. Much of the $k=20$ Gap reduction is already present at $k \in \{8,10\}$, allowing a smaller budget when model quality has higher priority.

\vspace{-4pt}

\vspace{-4pt}

\noindent\textit{\textbf{Finding 2.} The top-$k$ rankings overlap strongly, but the operators differ sharply. Subspace removal preserves model performance better than whole-head zeroing, which can increase GPT-2 language-modeling loss by up to $8\times$.}

\begin{figure}[!htbp]
\centering
\includegraphics[width=0.68\columnwidth]{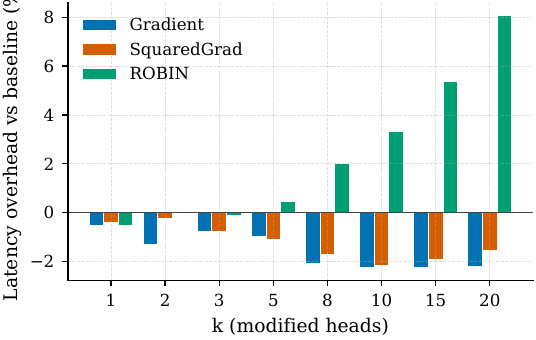}
\caption{Latency overhead on BERT as $k$ varies}
\label{fig:latency_overhead}
\end{figure}

\noindent
{\textbf{RQ3. How much inference cost does \textsc{ROBIN} add?}}\\
\noindent
\textsc{ROBIN} adds one matrix multiplication of shape $(d_h, r)$ per chosen head per forward pass. The resulting overhead at $k = 20$ ranges from $7$\% on the base-size models to $11$\% on the distilled models, because the same $k$ modifies a larger fraction of their attention heads (Table~\ref{tab:latency}). For the fixed batch size and sequence length we measured, overhead grows roughly linearly with $k$ (Fig.~\ref{fig:latency_overhead}).

\vspace{-4pt}
\begin{table}[!htbp]
\caption{Forward-pass latency (ms) at $k = 20$}
\label{tab:latency}
\centering
\scriptsize
\setlength{\tabcolsep}{2.8pt}
\begin{tabular}{lrrrr}
\toprule
Model & Baseline & Gradient & SquaredGrad & \textsc{ROBIN} \\
\midrule
DistilBERT  &  7.20 &  7.09\,($-1.5$\%) &  7.15\,($-0.7$\%) &  8.03\,($+11.4$\%) \\
BERT        & 12.04 & 11.81\,($-2.0$\%) & 11.82\,($-1.9$\%) & 12.98\,($+7.8$\%) \\
DistilGPT-2 &  8.35 &  8.27\,($-1.0$\%) &  8.29\,($-0.7$\%) &  9.25\,($+10.8$\%) \\
GPT-2       & 13.32 & 13.01\,($-2.3$\%) & 13.03\,($-2.2$\%) & 14.28\,($+7.2$\%) \\
\bottomrule
\end{tabular}\\[2pt]
{\scriptsize\itshape Note. Parentheses give change vs. baseline. Small negatives are measurement noise.}
\end{table}

\noindent\textit{\textbf{Finding 3.} For fixed batch size and sequence length, \textsc{ROBIN}'s measured latency overhead grows with $k$. The added operation has cost $O(BT k d_h r)$ per forward pass.}

\section{Discussion}

\textit{Relation to prior work.} SE fairness testing detects discriminatory behavior at the system boundary~\cite{galhotra2017fairness,udeshi2018automated}, while data-oriented repair changes training data or retrains the model~\cite{chakraborty2020fairway,zhang2021ignorance}. Transformer attribution ranks attention heads~\cite{michel2019sixteen,voita2019analyzing}, and recent search-based repair prunes selected heads~\cite{dasu2026attention}. Representation-level debiasing instead removes learned subspaces from whole-model representations~\cite{liang2020towards,ravfogel2020null}. \textsc{ROBIN} combines component localization with subspace removal restricted to selected head outputs. This white-box debugging contribution differs from whole-head pruning and does not claim a new fairness metric or causal explanation.

\textit{Operator vs.\ ranking.} Within each model, the three rankings overlap, with $15$ to $20$ shared heads in each top-$20$ list and Spearman correlations above $0.9$ between any two rankings. These overlaps make it unlikely that ranking differences alone explain the observed trade-off between fairness and model performance. This pattern is consistent with the per-head output carrying both bias and orthogonal components, since subspace removal targets the bias component while whole-head zeroing removes all components of the head output.

\begin{figure}[!htbp]
\centering
\begin{minipage}[t]{0.49\linewidth}
\centering
\includegraphics[width=\linewidth]{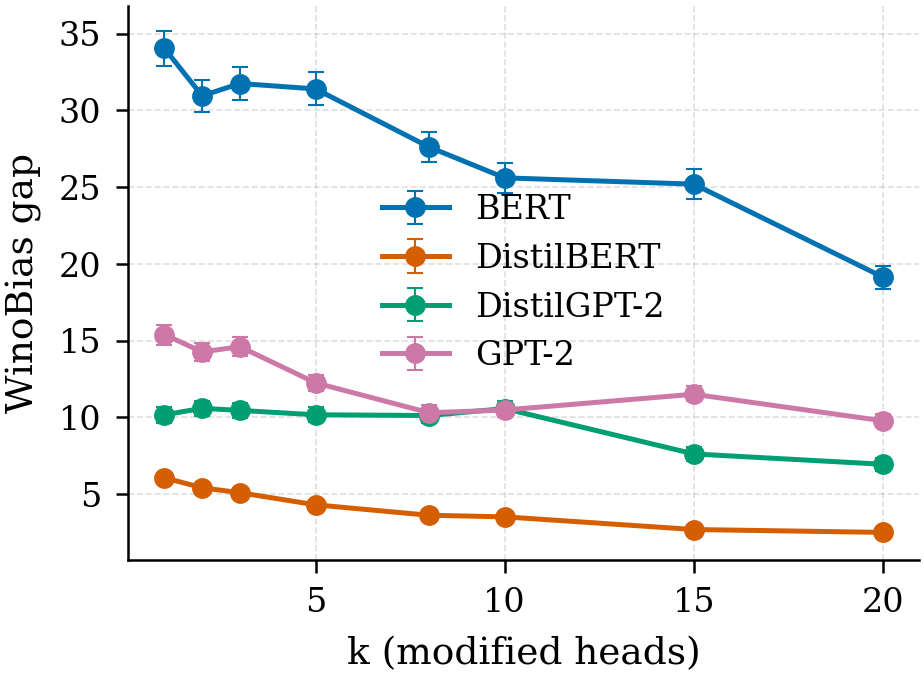}\\[-0.3em]
{\scriptsize (a) WinoBias gap}
\end{minipage}\hfill
\begin{minipage}[t]{0.49\linewidth}
\centering
\includegraphics[width=\linewidth]{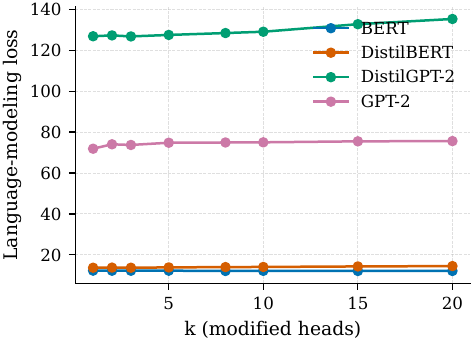}\\[-0.3em]
{\scriptsize (b) Language-modeling loss}
\end{minipage}
\caption{\textsc{ROBIN} across the four models as $k$ varies}
\label{fig:cross_model}
\end{figure}

\textit{Cross-model pattern.} Fig.~\ref{fig:cross_model} shows the same pattern across the models. Gap generally decreases with $k$, with small non-monotonic steps, and model performance stays close to the baseline.

\vspace{-1pt}

\textit{Scope and use.} \textsc{ROBIN} reduces a paired benchmark gap under our protocol. This result does not establish that the intervened models produce fairer outputs in realistic use, improve downstream-task fairness, or avoid new disparities outside the evaluated pairs. Gap and SS are proxy measures, while language-modeling loss and latency cover only broad model quality and runtime cost. These metrics do not determine the intervention's overall benefits and costs in deployment. \textsc{ROBIN} therefore provides candidate locations and a measurable intervention, not a causal explanation or a complete fairness assessment. A practitioner can choose $k$ from the measured trade-off and then validate model outputs in the intended context. This separation between localization and root-cause attribution matches standard SE debugging practice~\cite{wong2016survey}.

\textit{StereoSet.} StereoSet stereotype score does not move consistently across models and categories under our intervention. Some (model, category) combinations move toward neutrality and others away, without a consistent direction. Measurement-modeling work has documented validity concerns with StereoSet and similar contrastive benchmarks~\cite{blodgett2021stereotyping}, which is consistent with the noise we observe. We therefore base our main claim on WinoBias gap and language-modeling loss, where the fairness probes are paired by protected attribute and match the data used to estimate the per-head bias subspace. Generalizing to broader stereotype phenomena is a separate empirical question that this study does not answer.

\section{Threats to Validity}

Threats to \textit{internal validity} concern experimental choices that may affect the results. The fixed subspace dimension $r$ may change the trade-off between the WinoBias gap and language-modeling loss. We chose $r=4$ from a BERT pilot over $r \in \{1,2,4,8\}$ and held it fixed across models to avoid model-specific tuning. We use WinoBias for head ranking, subspace estimation, and the main gap evaluation. This reuse creates evaluation bias because the intervention may specialize to the same probe distribution used to construct it. StereoSet provides a secondary check, but its inconsistent results do not resolve this threat. A full evaluation requires disjoint data for ranking, subspace estimation, and fairness assessment. We also report means over three seeds without confidence intervals or statistical tests, so small differences between methods remain uncertain.

Threats to \textit{construct validity} concern whether the reported metrics capture fairness, model quality, and intervention cost. WinoBias covers gender bias in pronoun resolution, and its Gap metric assumes that the paired pronoun forms should receive similar scores. This assumption does not extend to contexts where a protected-attribute change is semantically relevant. StereoSet covers four stereotype categories, but neither benchmark establishes fairness in generated outputs or downstream tasks. WikiText-2 language-modeling loss does not capture all semantic output degradation, and latency does not capture all deployment costs. We therefore limit our claims to the evaluated probes, quality metric, and runtime setting.

Threats to \textit{external validity} concern generalization beyond the evaluated setting. The four models cover encoder, decoder, base, and distilled architectures, but not larger or instruction-tuned models. The latency results use one GPU and one batch shape, so the measured overhead may change under other hardware or implementations. We report the added cost as $O(BT k d_h r)$ per forward pass to clarify how the intervention scales.

\section{Conclusion and Future Work}
\label{sec:future}

Transformer language models can produce biased outputs whose internal contributors are difficult to localize. We propose \textsc{ROBIN}, a head-level fairness debugging method that selects attention heads and removes a small bias subspace from their outputs rather than zeroing them. In this pilot, \textsc{ROBIN} reduced the measured WinoBias gap while limiting language-modeling degradation. Future work will evaluate larger models, broader attributes, disjoint data, generated outputs, and downstream-task fairness. It will also study $r$ selection and subspace transfer.

\bibliographystyle{IEEEtranMN}
\bibliography{ref}

\end{document}